\documentclass[twocolumn,aps]{revtex4}
\usepackage{graphicx}
\usepackage{dcolumn}
\usepackage{bm}
\usepackage{latexsym}
\usepackage{graphics}                               
\usepackage{amssymb} 
\input isolatin1.sty

\date{\today}  
\begin{document}
\title{The crossover from 2D to 3D percolation\\
and its relationship to glass transition in thin films.\\
Theory and numerical simulations}

\author{Paul Sotta and Didier Long}
\affiliation{
Laboratoire de Physique des Solides\\
Université de Paris XI, Bât. 510\\
91405 Orsay Cedex, France}
\date{\today}
\begin{abstract}
We consider here the percolation problem in thin films, both
in the direction normal to the film and in the direction
parallel to the film. We thereby describe here
the cross-over between 2D and 3D percolation,
which we do on cubic and square lattices. The main relations are derived using
scaling and real space renormalisation arguments. They are
checked by numerical simulations, which also provide the numerical prefactors.
We calculate in particular the correlation length parallel to the film, the average mass and 
the mass distribution $n(m)$ of the clusters. In particular, we show that  
the latter is given
by a master function of $h^{-D+1/\sigma_{2}\nu_{3}}\vert p-p_{c}(h)\vert^{1/\sigma_{2}}
m$, where $h$ is the thickness of the film and $D,\nu_3,\sigma_2$ are tabulated 2D and 3D critical
exponents. $p_c(h)$ is the percolation threshold of the film which we
also calculate. These results are of interest in particular for 
describing  the glass transition in thin polymer films.
\end{abstract}

\pacs{Valid PACS appear here}
\maketitle

\section{\label{Intro}Introduction}
The building of macroscopic continuous objects (aggregates, or
clusters) by the random dispersion of particles (sites) or bonds
has been the subject of many studies over the past decades, and
has been formalised by various models of percolation
\cite{stauffer,stauff79,sahimi,essam}. The corresponding  issues
are essential for determining important macroscopic properties
such as the electric conductivity \cite{stauffer,sahimi}, the
visco-elastic behaviour \cite{degennes} of various systems, or
more generally mechanical properties \cite{obu}. For a review on
the latter aspect, see e.g. \cite{alexander}. The main quantities
of interest have been expressed as power laws, close to a critical
point, and the main results are summarised by a list of critical
exponents, which do not depend on the details of the models but
only on the spatial dimension. Due to the dependence of critical
phenomena on spatial dimension, the cross-over between 2D and 3D
behaviours has been the subject of various studies, e.g. in the
case of magnetic phenomena \cite{domb}, for describing the phase
transition behavior of thin magnetic films for instance. For
studying the conductivity of thin composite systems, Clerc et al
considered the cross-over between 2D and 3D percolation and
calculated the percolation threshold as a function of the
thickness of the film, using finite scaling arguments
\cite{clerc}. This issue has also been considered by Vicsek and
Kertész \cite{vicsek}.

Another field where percolation appears to be a key concept is the
glass transition in supercooled liquids. The most prominent
feature of this phenomena is a dramatic increase of the viscosity
when cooling such liquids. Though very steep, this increase is
continuous and does not appear to involve a phase transition. For
a review on the glass transition see e.g. \cite{ediger,science}.
To account for experiments which have demonstrated the
heterogeneous nature of the dynamics when approaching the glass
transition \cite{spiess1,spiess95,spiess2,ediger1,ediger12} (see
e.g. \cite{ediger5,sillescu,richert} for review), it has been
proposed recently \cite{long1,long2} that the glass transition
corresponds to the percolation of domains with relaxation times
larger or equal to the arbitrary time $\tau_g$ set for defining
the glass transition. According to this model, the slow domains
correspond to denser regions resulting from density fluctuations
of Gaussian statistics. These slow domains, of typical size 2 to 4
nm, coexist with faster domains with relaxation times order of
magnitude shorter \cite{sillescu,ediger5,richert,long1,long2}.
This model, aimed at explaining experimental results regarding the
heterogeneous nature of the dynamics close to the glass
transition, allows also to explain the shift of the glass
transition temperature $T_g$ of thin polymer films, a few tens of
nanometer thick. Indeed, it has been demonstrated over the past
ten years that the dynamical behaviour of thin polymer films is
very different from that of the same polymer in the bulk. It is
now well established that films with weak interactions with their
susbstrate (or freely suspended films) display a reduction of
$T_g$, of about $20$ $K$ for films $10$ $nm$ thick
\cite{jones1,jones2}. On the contrary films with strong
interactions with their substrate display an increase of $T_g$ by
as much as $60 K$ for films $10$ $nm$ thick
\cite{wallace1,grohens,grohens2}. For review of the corresponding
issues see e.g. \cite{dalnoki,forrest3}. In the model of Long and
co-workers \cite{long1,long2,long3}, these effects result from
percolation mechanisms. According to this model, the glass
transition is controlled by the percolation of small subunits of
relaxation time $\tau_g$. Therefore, the shifts in $T_g$ in a film are
closely related to the differences in the percolation properties in
a film (i.e. in a system of small, finite thickness) with respect
to a bulk system. In the case of a thin suspended film,
percolating in the direction parallel to the film requires a
larger fraction of these slow subunits, and thereby takes place at
a temperature lower than the bulk $T_g$.

In the case of a strongly interacting film, the glass transition
occurs when aggregates of slow subunits have a diameter comparable
to the thickness of the film, due to different boundary
conditions. As a consequence, in both cases the glass transition
corresponds to the temperature at which the correlation length of
the 3D percolation problem is equal (or comparable) to the
thickness of the film, in one case (weak interactions) above the 3D 
percolation threshold, and in the other one (strong interactions)
below the 3D percolation threshold. In the case of intermediate
interactions between the monomers and the substrate
\cite{nealey1,green}, as described in \cite{long3}, the glass
transition corresponds to a lateral extension of a slow aggregate
larger than the film thickness, and such that the number of
monomers of this aggregate in contact with the substrate is large
enough so that the time for this aggregate to desorb is equal to
$\tau_g$. Describing all these cases -weak interactions, strong
interactions, intermediate interactions- requires thereby a
precise understanding of the cross-over between 2D and 3D
percolation \cite{long1,long2}. This is the aim of this paper. A
better description of this cross-over should also be useful for
describing the mechanical properties of films made of composite
materials, or for the conductivity properties of thin films
\cite{stauffer,clerc}.

The questions which we address here stem from our modeling of the
glass transition in thin films. When considering a thin suspended
film, we argued \cite{long1,long2} that the glass transition
corresponds to the percolation in the direction parallel to the
film. Then, we will consider this problem here, in films of
dimensionless thickness $h$, expressed in units of the size $\xi
\sim$ 2-4 nm
 of the dynamical heterogeneities \cite{spiess2,long2}. Then, one of the
questions is: what is the percolation threshold $p_c(h)$ knowing
the 3D percolation threshold $p^{3D}_c$? How does the lateral
extension of the aggregates increase when approaching this
threshold? When considering films strongly attached to their
substrate, we argued that the glass transition corresponds to the
appearence of a fraction of order $1$ of monomers from one
interface connected to the opposite interface by continuous path
of slow subunits. We thus study the correlation function
between sites on one interface and sites on the other, to
determine precisely how the connected fraction evolves with the
fraction $p$ of occupied sites. Finally, to consider the crossover
between these two regimes, we need to know the size of the
aggregates, the number of monomers of one aggregate in contact
with the interacting substrate, and the distribution of mass of
the aggregates. 

The present work is based on the idea that a
system of finite thickness $h$ may be renormalised to a 2D system,
by changing the 2D percolation threshold $p_{c}^{2D}$ into a
renormalised value $p_{c}(h)$. A first step will be therefore to
determine $p_{c}(h)$. We describe a real-space renormalisation
procedure to transform the quasi-2D problem of a film of finite
thickness into a 2D problem. This allows for calculating various
quantities of interest, as a function of known 2D and 3D critical
exponents and of a renormalised occupation probability. The
corresponding results are exact at the level of scaling laws. To
check these predictions, and to obtain the values of the
pre-factors, we then perform numerical simulations.
The paper is organized as follows. In section {\bf II.A}, we introduce 
some classical definitions, notations and we recall some basic
concepts and results regarding  percolation theory, either in 2D or in 3D. 
Specifically we recall the
various quantities which have a critical behaviour at the
percolation threshold. In section {\bf II.B}, we introduce some
classical finite size scaling arguments regarding the definition of
the percolation threshold on a finite
system. Then in section {\bf II.C}, we show how the percolation
problem in a film of finite thickness can be mapped on
a 2D system with a new percolation threshold $p_c(h)$. Various
scaling laws are derived, regarding $p_c(h)$, the mass of the
aggregates, the connectivity between both interfaces.
In section {\bf III} we describe the numerical Monte-Carlo algorithm used for
our simulations. Finally in sections {\bf IV}, {\bf V} and {\bf VI}
we discuss the results of our numerical simulations, in connection 
with the various scaling laws derived in {\bf II.C}.

\section{\label{backgrnd}Background on percolation}
\subsection{\label{critic}Critical behaviour at $p=p_c$}
Let us recall some basics about 2D or 3D percolation
which will be useful in the following.
For a review of the definitions and results presented
here see e.g. \cite{stauffer}. The control parameter in site
percolation problems is the site occupation probability $p$. The
percolation threshold is the critical value $p_c$ at which an
infinite cluster appears. The value $p_c$ is not universal. It
depends on the particular type of lattice which is considered.
Values of the percolation thresholds in various cases are
summarized in Table \ref{tab:table1}. In the following,
the various exponents are generically designed by greek letters.
When some relations are valid only using a 2D or a 3D exponent,
the corresponding exponents are written with a subscript
2 or 3 according to the dimension.

\begin{table}
\caption{\label{tab:table1}The values of the percolation
thresholds in various cases}
\begin{ruledtabular}
\begin{tabular}{lcr}
Lattice&Site&Bond\\
\hline
square (2D) & 0.592746 & 0.50\\
simple cubic & 0.3117 & 0.2492\\
FCC & 0.198 & 0.119\\
\end{tabular}
\end{ruledtabular}
\end{table}
For the sake of definiteness and when it is necessary to be
specific, we will consider in this paper percolation on cubic
lattices (3D or in films), and on square lattices for 2D systems.
Only the prefactors depend on this choice, the exponents of the
various scaling laws that we derive do not. We call $n(m)$ the
number of clusters of mass $m$ (the mass $m$ of a cluster is the
number of sites which belong to it) per lattice site (i.e. the
total number of clusters of size $m$ divided by the volume of the
system), at a given occupation probability $p$. The quantity
$m\,n(m)$ therefore equals the probability for a given site to
belong to a cluster of mass $m$. The occupation probability is
thus $p=\sum m \, n(m)$. The distribution of cluster masses is
represented by a function of the form:
\begin{equation}
n(m) \approx m^{-\tau} \,f(m/m_{\zeta}) \label{eq:2bis}
\end{equation}
where $f(m/m_\zeta)$ behaves like
\begin{equation}
f(m/m_\zeta) \approx  \,e^{-m/m_{\zeta}} \label{eq:2}
\end{equation}
for  $m \gg m_\zeta$ and has a finite value at zero. For $m$
smaller than the typical value $m_{\zeta}$, the power law
dominates. For $m$ larger than $m_{\zeta}$, the exponential
dominates, which means that clusters with a mass larger than
$m_{\zeta}$ are exponentially rare. $m_{\zeta}$ diverges at $p_c$
with a critical exponent denoted by $1/\sigma$:
\begin{equation}
m_{\zeta} \approx \vert p-p_c \vert ^{-1/\sigma}\label{eq:2ter}
\end{equation}
In 2D, the exponent $\sigma_{2} =36/91=0.3956$ (or $1/\sigma_{2}
=2.5278$) and the exponent $\tau_{2} =187/91=2.055$ (see
Table~\ref{tab:table1}).

On the other hand, a mean cluster mass may be defined for $p<p_c$
as the second moment:
\begin{equation}
M = \frac{\sum m^{2}n(m)}{\sum m \, n(m)} \label{eq:1}
\end{equation}
The mean cluster mass $M$ diverges at the threshold $p=p_c$ with
another exponent $\gamma$:
\begin{equation}
M \approx \vert p-p_c \vert ^{-\gamma} \label{eq:3}
\end{equation}
In 2D: $\gamma_2 = 43/18 = 2.389$.\\
Since $\sum m \, n(m) \equiv p$ (or equivalently, since $\tau
>2$), the divergence is contained in the numerator in
eq.~(\ref{eq:1}), which reads:
\begin{equation}
M \approx \int m^{2-\tau}f \left( \frac{m}{m_\zeta} \right) dm
\propto m_{\zeta}^{3-\tau} \label{eq:4-1}
\end{equation}
Note that $M$ grows with $|p-p_c|$ more slowly than $m_\zeta$. This is
because $M$ is an average mass (second moment of the cluster number $n(m)$), 
whereas $m_\zeta$ is the cut-off mass of the distribution. 
Combining eq.~(\ref{eq:2ter}),~(\ref{eq:3})
and ~(\ref{eq:4-1}), one finds:
\begin{equation}
\gamma = \frac{3-\tau}{\sigma}  \label{eq:4}
\end{equation}
Note that, at $p=p_c$, the distribution $n(m)$ reduces to $n(m)
\approx m^{-\tau}$.

We consider now the spatial extension of the clusters. One can
define the correlation function (or connectivity function) $g(r)$
as  the probability that a site at the distance $r$ from an
occupied site is also occupied and belongs to the same cluster.
The correlation length may be defined from $g(r)$ as:
\begin{equation}
\zeta^2  =
 \frac{\sum_r r^{2} g(r) }{\sum_r g(r)} \label{eq:4-6}
\end{equation}
For $p<p_c$,
$g(r)$ is represented by a function of the form:
\begin{equation}
g(r) \approx r^{-(d-2+\eta)}G(r/\zeta) \label{eq:4g}
\end{equation}
where $d$ denotes the dimensionality of space, and $\eta$ is the
anomalous exponent of the correlation function \cite{stauffer}.
For $r \ll \zeta$, the power law dominates. This means that at
short scale, the system is insensitive to the presence of a
cut-off length at a larger scale. At large scale ($r \gg \zeta$),
the function $G$ decays exponentially $G(r/\zeta) \approx
\exp(-r/\zeta)$. The correlation length $\zeta$, which is
proportional to some typical cluster diameter, diverges at the
threshold as:
\begin{equation}
\zeta \approx \vert p-p_c \vert ^{-\nu} \label{eq:4bis}
\end{equation}
where, in 2D: $\nu_2 = 4/3$, and in 3D: $\nu_3  \approx 0.88$.

The radius of a cluster of mass $m$ may be defined as:
\begin{equation}
2R^2_m = \sum \frac{|r_{ij}|^{2}}{m^2} \label{eq:4-5}
\end{equation}
where $|r_{ij}|^{2}$ is the squared distance between two occupied
sites which belong to the same cluster of mass $m$, and where the
sum is then averaged other all clusters of mass $m$. According to
the definitions, it is easy to check that one has:
\begin{equation}
\zeta^2 = \frac{\sum R^{2}_m m^2 n(m)}{\sum m^2 \, n(m)}
\label{eq:4-6}
\end{equation}
The correlation function $g(r)$ is related to the structure of the
clusters, i.e. to the distribution of the mass inside a cluster. In 3-D, let
us draw a volume of size $h$ (of volume $h^3$) within a large
cluster, such that $h$ is small with respect to $\zeta$. The number
of sites within this volume which belong to this cluster is:
\begin{equation}
m=\int_{0}^{h}g(r)r^{2}dr  \label{eq:4-2}
\end{equation}
And for $h<\zeta$:
\begin{equation}
m= \int_{0}^{h} r^{1-\eta}dr\approx h^{2-\eta}
\label{eq:4-3}
\end{equation}
In 3D, the exponent $\eta$ is relatively small: $\eta=-0.068$.
Setting $h=\zeta$ in eq.~(\ref{eq:4-3}), one obtains the average
mass $M$, as defined in eq.~(\ref{eq:3}): $M \approx
\zeta^{2-\eta}$, which gives another relation between critical
exponents:
\begin{equation}
(2-\eta)\nu=\gamma \label{eq:4-4}
\end{equation}

We consider now the distribution of radius of the clusters for $p
\le p_c$. The radius $R_m$ of a cluster is related to its mass $m$
by a scaling law:
\begin{equation}
m \approx R^D_m \label{eq:5}
\end{equation}
However, the exponent (or apparent exponent) $D$ is not the same
below, at and above $p_c$ \cite{stauffer}. For $p = p_c$, the 3D
critical exponent is $D_3 \approx 2.53$, and for $p < p_c$, the
scaling relation, i.e. the fractal exponent, depends on the
considered radius. Thus, more generally, it is assumed that the
relation between the mass and the spatial extension is of the form
\cite{stauffer}:
\begin{equation}
R_m \approx m^{1/D} f[ (p-p_{c}) m^{\sigma}]
 \label{eq:15}
\end{equation}
where the function $f$ has to be determined numerically and
satisfies the following properties: as $m \ll m_{\zeta} \approx
\vert p-p_{c} \vert ^{-1/\sigma}$ (where $m_{\zeta}$ has been
defined in eq.~(\ref{eq:2ter})), i.e. as $x= \vert p-p_{c} \vert
m^{\sigma} \ll 1$, $f(x)$ approaches a constant value. For $x \gg
1$, $f(x) \sim x^{1/D'-1/D}$, where $D'$ is the fractal exponent
of large clusters for $p<p_c$. Away from $p_c$, $m_{\zeta} \approx
\vert p-p_{c} \vert ^{-1/\sigma}$ is finite, and defines a
crossover from the behaviour $R_{m} \approx m^{1/D}$ for
$m<m_{\zeta}$ to another power law $R_{m} \approx m^{1/D'}$ for
$m>m_{\zeta}$. At $p_c$, $m_{\zeta}$ diverges. The exponent $1/D$
in eq.~(\ref{eq:15}) is therefore the one which is measured at
$p_c$ (or very close to $p_c$). This exponent takes the values
$1/D_{3}=1/2.53=0.39526$ in 3D and $1/D_{2}=48/91=0.5275$ in 2D.
For clusters with a mass larger than $m_{\zeta}$, at $p<p_c$, the
value of the fractal exponent is smaller: $D_3 = 2$. This means
that for $p < p_c$, large clusters are more tenuous than small
ones, while they are self-similar whatever the size at $p_c$. 
The above discussion means that the crossover
takes place on a spatial scale $\zeta$. For small clusters with
$R_m < \zeta$ (or for small masses $m < m_\zeta$), the value of
the critical exponent is that at $p = p_c$ ($D \approx 2.53$ in 3D), 
and for larger clusters the exponent is that at $p<p_c$ ($D = 2$ in 
3D). The cross-over
between these two behaviours takes place at $R_m = \zeta$, corresponding to
the mass $m_\zeta = \zeta^D$, where $D$ is the critical exponent
at $p_c$. Note that the cross-over goes to larger scales when $p$
tends to $p_c$ and there is a single exponent at $p_c$. All this means
that at short scales (i.e. at scales significantly shorter than
the correlation length $\zeta$), the system does not know whether
it percolates at larger scales or not.

Combining eqs.~(\ref{eq:2ter}),~(\ref{eq:4bis}) and~(\ref{eq:5}),
one gets another relation between critical exponents:
\begin{equation}
\frac{1}{\sigma}=\nu D \label{eq:6}
\end{equation}
where $D$ has the value at $p=p_c$ here.
\begin{table}
\caption{\label{tab:exponents}The values of some of the critical
exponents, which will be used in the following}
\begin{ruledtabular}
\begin{tabular}{lcr}
Exponent&$d=2$&$d=3$\\
\hline
$\tau$ & 187/91=2.055 & 2.18\\
$\sigma$ & 36/91=0.3956 & 0.45\\
$\gamma$ & 43/18=2.389 & 1.80\\
$\nu$ & 4/3 & 0.88\\
$1/D (p=p_{c})$ & 49/91=0.5275 & 1/2.53=0.3952\\
$D (p<p_{c})$ & 1.56 & 2\\
$\eta$ & 0.2084 & -0.068\\
\end{tabular}
\end{ruledtabular}
\end{table}

\subsection{\label{finite}Percolation in a finite system}
The problem in numerical simulations is: how to extract the value
of the percolation threshold (which is defined at infinite size)
from simulations done in boxes of finite size ? Percolation in a
system of size $L$ is defined by the appearance of one cluster of
radius $L$. The probability that a system of linear size $L$ (volume $L^3$) 
percolates
at an occupation probability $p$ is of the form:
\begin{equation}
\Phi_{L}(p) \approx f \left( L \left( p_{c}-p \right)^{\nu}\right)
\label{eq:10}
\end{equation}
where the exponent $\nu$ has been introduced in
eq.~(\ref{eq:4bis}). Eq.~(\ref{eq:10}) means that the probability
to percolate in a system of size $L$ is identical to the
probability of appearance of a cluster of radius $L$ in an
infinite system, or, equivalently, that a system of size $L$
percolates when the correlation length $\zeta(p) =L$ (where $\zeta$ is
defined e.g. in eq.~(\ref{eq:4bis})). According to
eq.~(\ref{eq:10}), the curves $ \Phi_{L}(p)$ become steeper as $L$
increases. We call $p_{c}(L)$ the occupation probability
corresponding to a given probability to percolate $P$. It follows
from eq.~(\ref{eq:10}) that in a system of size $L$, $p_{c}(L)$ varies as:
\begin{equation}
\vert p_{c}(L)-p_{c} \vert \approx L^{-1/\nu}
\label{eq:11}
\end{equation}
Eq.~(\ref{eq:11}) will be used in the following to determine
percolation thresholds $p_c$ (extrapolated at infinite size) from
simulations in boxes of various finite sizes. The percolation 
threshold $p_c(n)$ will be given by the number 
giving the best regression with exponent $-1/\nu$ (according to
eq.~(\ref{eq:11})).

\subsection{\label{rescale}Rescaling a system of finite thickness $h$}
\begin{figure}
\includegraphics{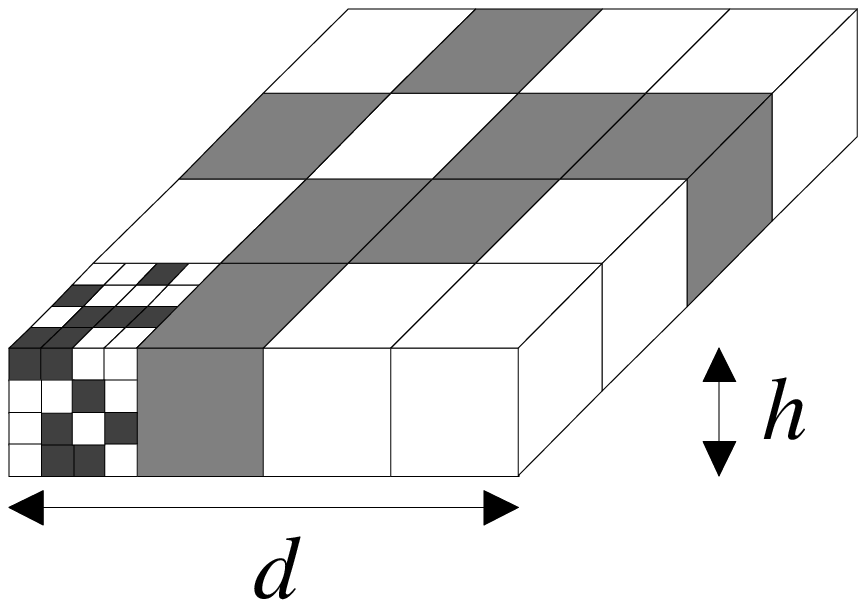}
\caption{\label{fig1} Schematics illustrating the renormalization
of the percolation problem in a film of thickness $h$ to a two
dimensional system. }
\end{figure}

Consider a thin film  with mesh size unity,
probability of site occupation $p$ and of thickness $h$. 
The probability of occupation $p$ may be larger
or smaller than $p^{3D}_c$ and is such that the corresponding $3D$
correlation length $\zeta_{3D}$ is larger than the thickness $h$
of the film, that is:
\begin{equation}
p^-_c(h) \lesssim p \lesssim p^+_c(h)
\end{equation}
The quantities $p^+_c(h)$ and $p^-_c(h)$ correspond to a
correlation length for the 3D percolation problem comparable to
the thickness of the film. $p^-_c$ is smaller than the 
3D percolation threshold $p^{3D}_c$, and $p^+_c$ is larger than $p^{3D}_c$. 
We aim now at calculating the percolation threshold of the film, $p_c(h)$. 
Note that we consider here percolation in the direction parallel 
to the film. 
By a coarse graining procedure, this
problem can be transformed into a $2D$ problem. The corresponding
coarse graining function is denoted by $\Phi$. After one
iteration, the probability of occupation is transformed into
$\Phi(p)$. $p^{3D}_c$ is the fixed point of $\Phi$. Let us define
$\kappa > 1$ as the scale factor of the renormalisation procedure.
$N$ is a number such that
\begin{equation}
\kappa^N = h
\end{equation}
which means that the film is transformed into a 2D squared lattice
after $N$ iterations. If the initial film is at the percolation
threshold, one requires that after these $N$ iterations the film
is transformed into a 2D lattice close to the 2D percolation threshold,
that is $p=p_c(h)$ is transformed into $\Phi^N(p_c(h))$ such that
\begin{equation}
\Phi^{(N)}(p_c(h)) -p^{3D}_c \sim p_c^{2D} - p_c^{3D}
\label{eq:13-5}
\end{equation}
One has then
\begin{equation}
\lambda^N(p_c(h) -p^{3D}_c) = \alpha
\end{equation}
where $\alpha$ is a positive number comparable to $p_c^{2D} -
p_c^{3D}$ and where
\begin{equation}
\lambda = \frac{d \Phi}{dp}(p=p^{3D}_c)
\end{equation}
Therefore
\begin{equation}
p_c(h) \approx   p^{3D}_c + \alpha \lambda^{-N}  \approx  p_c^{3D} + \alpha
h^{-1/\nu_3} \label{eq:13-6}
\end{equation}
where $\nu_3$ is the 3D critical exponent for the correlation
length, which satisfies the relation:
\begin{equation}
\nu_3 = \frac{\log\kappa}{\log\lambda}
\end{equation}
Note that the film percolation threshold $p_c(h)$ is larger than $p^{3D}_c$ -percolating in thin
films is more difficult than in the bulk- and that at $p_c(h)$ the
3D correlation length is comparable to $h$. In the following we 
take  $p^+_c(h) =  p_c(h)$. Let us consider now a situation where the
probability of site occupation in the film is $p<p_c(h)$. By the
same argument, $p$ is transformed into $\Phi^{(N)}(p)$ with
\begin{equation}
\Phi^{(N)}(p) -p^{3D}_c \approx \lambda^N(p -p^{3D}_c)
\end{equation}
This procedure is illustrated in Fig.~\ref{fig1}.
One has then
\begin{equation}
\Phi^{(N)}(p_c(h)) -\Phi^{(N)}(p) \approx \lambda^N(p_c(h) -p) \approx 
h^{1/\nu_3}(p_c(h) -p)
\end{equation}
The quasi-2D problem of the film with probability of occupation
$p$ on a lattice of mesh size unity is then mapped on a 2D lattice
with probability of occupation $\Phi^{(N)}(p)$ and mesh size $h$.
The average mass $M'(h)$ of the aggregates, expressed in numbers
of supersites of size $h$, is therefore:
\begin{eqnarray}
M'(h) \approx M'_0 |\Phi^{(N)}(p_c(h)) -\Phi^{(N)}(p)|^{-\gamma_2}
\\ 
\approx M_{0}' \left( h^{1/\nu_{3}} \vert p-p_{c}(h)
\vert \right)^{-\gamma_{2}}
\label{eq:14-0}
\end{eqnarray}
where $M'_0$ is a prefactor of order unity. Note that the above
result insures that for $p = p^-_c(h)$, i.e. when the $3D$
correlation length is equal to the thickness of the film, the
typical mass of the aggregates on the scale $h$ is of order one as
it should be. Note indeed that $p^-_c(h)$ is given by a relation 
\begin{equation}
p^-_c(h) = p^{3D}_c - \alpha'h^{-1/\nu_3}
\label{eq:pcm}
\end{equation} 
where $\alpha'$ is a positive number of order $1$. 
Let us study now the structure of the clusters in
the film, below but close to the percolation threshold $p_{c}(h)$,
more precisely in the situation $h \ll \zeta_{3D}$. In this case,
the lateral extension of the dominating cluster is large with
respect to $h$. The 2D correlation length is, expressed in the
equivalent renormalised 2D system:
\begin{equation}
\zeta'_{\|} \approx \vert \Phi^{(N)}(p)-p_{c}^{2D} \vert^{-\nu_{2}} \approx  \left(
h^{1/\nu_{3}} \vert p-p_{c}(h) \vert \right)^{-\nu_{2}}
 \label{eq:14bis}
\end{equation}
Coming back to the original system, in units of elementary sites:
\begin{equation}
\zeta_{\|} \approx h \left( h^{1/\nu_{3}} \vert p-p_{c}(h) \vert
\right)^{-\nu_{2}}
 \label{eq:14ter}
\end{equation}
Here $\zeta_{\|}$ is the correlation length in the direction
parallel to the film and provides the dominant lateral extension
of the clusters. Of course this result is meaningful only when the
occupation probability $p$ is such that the 3D correlation length
is larger than the thickness of the film. In units of the
elementary sites of the initial film, $M_{0}'$ should then be
replaced by the average mass within a $h$-cube. This is done using
eq.~(\ref{eq:4-3}): $M_{0}'= M_{0}h^{2-\eta}$ (because in all the
interval studied here $h<\zeta_{3D}$). $M_0$ is a number of 
order unity. Finally, the average mass
per cluster is, in units of elementary sites:
\begin{equation}
M \approx M_0 h^{2-\eta_{3}} \left( h^{1/\nu_{3}} \vert p-p_{c}(h)
\vert \right)^{-\gamma_{2}}
 \label{eq:15bis}
\end{equation}

Let us consider now the distribution of clusters $n(m)$. It obeys
a scaling relation similar to eq.~(\ref{eq:15}) (see also
eq.~(\ref{eq:2bis})). In the equivalent renormalised 2D system,
the number of clusters of mass $m'$ (in units of $h$-cubes) is
(see eqs.~(\ref{eq:2bis}) and ~(\ref{eq:2ter})):
\begin{equation}
n(m') \approx  m'^{-\tau_{2}}g_{2}\left(\frac{m'}{m_{\zeta}'}\right)
 \label{eq:15ter}
\end{equation}
with:
\begin{equation}
m_{\zeta}' \approx \vert
\Phi^{(N)}(p)-p_c^{2D}\vert^{-1/\sigma_2} \approx h^{-1/\nu_3\sigma_2}\vert
p-p_c(h)\vert^{-1/\sigma_2}
 \label{eq:15-4}
\end{equation}
$g_{2}$ is the 2D function as defined by equation (1), and has to
be determined numerically. $\tau_{2}=187/91=2.055$ and
$1/\sigma_{2}=91/36=2.5278$ are the 2D values of the exponents
(see Table~\ref{tab:table1}). The expressions for the masses
$M'$ and $M$ can be obtained in a different way as that used
above, which we consider now. From the mass distribution $n(m')$, one deduces the
relation
\begin{equation}
M' \approx m_{\zeta}'^{3-\tau_2}
 \label{eq:15-5}
\end{equation}
Using the same argument as that used to obtain equation
~(\ref{eq:15bis}), one obtains
\begin{equation}
M \approx h^{2-\eta_3} m_{\zeta}'^{3-\tau_2}
 \label{eq:15-6}
\end{equation}
Then, from equation ~(\ref{eq:15-4}), the average mass $M'$ on scale
$h$ can be written as
\begin{equation}
M' \approx
h^{-(3-\tau_2)/(\nu_3\sigma_2)} |p-p_{c}(h)|^{-(3-\tau_2)/\sigma_2}
 \label{eq:15-7} 
\end{equation}
Finally, using the relation $(3-\tau_2)/\sigma_2 = \gamma_2$, one recovers
\begin{equation}
M' \approx
h^{-\gamma_2/\nu_3} |p-p_{c}(h)|^{-\gamma_2}
 \label{eq:15-8}
\end{equation}
which is equation ~(\ref{eq:14-0}). To come back to the original
film of thickness $h$, we need now to obtain the distribution
$n(m)$ of the mass $m$ of the aggregates on the scale of
elementary sites. When the system is rescaled by a factor $\kappa$
(in 3D), the correlation length is rescaled as $\zeta \rightarrow
\frac{\zeta}{\kappa}$. Since the relation
$m_{\zeta}=\zeta^{-1/(\sigma\nu)}$ must be preserved, $m_{\zeta}$
is thus rescaled as:
\begin{equation}
m_{\zeta} \rightarrow \frac{m_{\zeta}}{\lambda^{1/(\sigma_3
\nu_3)}}
 \label{eq:15-2}
\end{equation}
With $1/(\sigma \nu)=D$ (see eq.~(\ref{eq:6})), this leads to the
rescaling $m_\zeta=h^{D_3}m'_\zeta$, where $D_3$ is here the 3D
value of the exponent at $p=p_c$. Thus, we have to consider a
scaling relation of the type:
\begin{equation}
n(m) \approx h^{\omega}m^{-\tau_2} g_{2D} \left[
h^{-D_3+1/(\sigma_{2}\nu_{3})}\vert
p-p_{c}(h)\vert^{1/\sigma_{2}} m\right]
 \label{eq:16-0}
\end{equation}
or equivalently
\begin{equation}
n(m) \approx h^{\omega}m^{-\tau_2} g_{2D}
(h^{-D_3}\frac{m}{m'_\zeta})
 \label{eq:16-1}
\end{equation}
where $\omega$ can be determined by writing:
\begin{equation}
M = \int m^2 n(m) dm  \approx
h^{\omega+D_3(3-\tau_2)}m'^{3-\tau_2}_\zeta
 \label{eq:16-2}
\end{equation}
Comparing with eq.~(\ref{eq:15-6}), one obtains the equation
$\omega+D_3(3-\tau_2) = 2-\eta_3$. Thereby the distribution of
cluster masses is expected to be described by equation
~(\ref{eq:16-0}) or equation ~(\ref{eq:16-1}), the value of the
exponent $\omega$ being given by
\begin{equation}
\omega = 2-\eta_3-D_{3}(3-\tau_2)
\end{equation}
Thus, the quantity $h^{-\omega}m^{\tau_{2}}n(m)$ should give a
master curve when plotted versus the quantity
$h^{-D_3+1/(\sigma_{2}\nu_{3})}\vert
p-p_{c}(h)\vert^{1/\sigma_{2}} m$. The exponent $\omega$ has the
value $\omega=-0.323$. In  the following,  
we consider numerical simulations regarding the relations derived 
in this section.

\section{\label{algor}Description of the system and the
algorithm}

We compute the site percolation threshold in a square (2D) or
simple cubic (3D) lattice of variable linear dimensions. Unit
sites are occupied at random with a probability $p$. An improved
random number generator (uniform deviate provided in numerical
recipes in Fortran), is used, which, thanks to a shuffling
procedure of the output, is known to provide perfect random
numbers, within the limits of the floating point precision.
Indeed, the random generator has to be called a number of times
equal to the volume of the system times the number of sample
systems used to average the data (that is, at most a few
$10^{11}$), which precludes the use of a standard, low quality
random number generator.

The algorithm used to determine percolation thresholds is itself
based on a renormalization procedure, which is either two- or
three-dimensional. For instance, at a given step in the 3D
procedure, $d$-cubes of linear size $d=2^m$ (of volume $V=d^3$)
are generated. The faces of the cube are occupied by atomes
belonging to various clusters. Then, 8 such $d$-cubes are gathered
to form one cube of linear size $2d=2^{m+1}$ (volume $8V$), in
which the clusters from all 8 elementary cubes are connected
together and re-numbered. A Hoshen-Kopelman algorithm is used to
renumber the clusters \cite{hoshen}. Then only the new clusters
connected to one external face of the new cube at least are
considered for the next step. The initial step consists in
connecting together 8 unit sites of the lattice. Iterating the 3D
procedure $m$ times gives a system (a cube) of size $2^m$. Then, a
film of thickness $h$ and lateral dimension $d=h^p$ may be
generated by iterating $p$ times a 2D procedure similar to the 3D
one. The system percolates when there exists at least one cluster
connecting the 6 external faces. The probability to percolate
$P(p)$ is averaged over a number of different configurations
(typically 50 to 500). Using the renormalization procedure, the
curves $P(p)$ for all values of $d$ (of the form $d = 2^j$) are
obtained in a single run. Simulations were run on Pentium IV
Personal Computers operating at 1.4 GHz with 512 Mo RAM or on a
DecAlpha Work Station.

The algorithm was validated by studying the well-known cases of 2D
(resp. 3D) square (resp. simple cubic) systems, in order to verify
that percolation thresholds are obtained with a satisfactory
precision. The probability to percolate $P$ is measured as a
function of the occupation probability $p$. In two dimensions, the
curves $P(p)$ obtained for various sizes of the system  (from $d =
2^{3} = 8$ to $d = 2^{16} = 65536$, which corresponds to a surface
$d^2$ from 64 to $4.295\times 10^9$) become steeper when $d$
increases, in agreement with eq.~(\ref{eq:10}). We find here that all curves
intersect at a fixed point which corresponds to an occupation
probability $p_c^{2D} \approx 0.5927$ and a probability to percolate
$P(p) \approx 0.327$. The value $p_c$ is in agreement with that
given in \cite{stauffer}. The curves $P(p)$ are well 
fitted (at least in the vicinity of the percolation threshold) by
a function of the form:
\begin{equation}
P(p)=\frac{1}{2} \left( 1+\tanh \left(\frac{p-p_{c}(d)}{q (d)}
\right) \right) \label{eq:17}
\end{equation}
in which both adjustable parameters $p_c(d)$ and $q(d)$ depend on
the size $d$ of the system. The form of this function is purely
heuristic, as it merely provides a unambiguous way to extract the
value of the threshold for each value of $d$.
When the quantity $\log(p_c(d) - p_{c}^{2D})$ is plotted as a
function of $\log d$, a straight line with a slope $-0.75 \pm
0.01$ is obtained. This is in concordance with eq.~(\ref{eq:11}),
with  the theoretical value $-1/\nu_{2}=-3/4$. The quantity
$q(d)$, plotted as a function of $d$ in logarithmic scale does
also give the power law behaviour $q(d)\approx d^{-1/\nu_{2}}$, in
agreement with eq.~(\ref{eq:10}).
In 3D as well as in 2D, the curves $P(p)$ obtained for various
sizes of the system (from $d = 8$ to $d = 2^9 = 512$, i.e. a
volume 512 to $1.342\times 10^8$) strongly depend on $d$. All
curves intersect at the fixed point $p_c^{3D} = 0.3117$ and a
probability to percolate $P(p) \approx 0.073$. The value
$p_c^{3D}$ is in agreement with that given in 
\cite{stauffer}. The curves $P(p)$ are fitted by a function of the
same form as in the 2D case (eq.~(\ref{eq:17})).
In 3D as well, the quantity $p_{c}(d)-p_{c}^{3D}$ varies as
$d^{-1/\nu_{3}}$, with the expected 3D value of the exponent
$1/\nu_{3} = 1/0.88=1.1364$. Note that the prefactor in this curve
(as well as in the 2D curve) is arbitrary, as it depends on the
probability level $P$ at which $p_{c}(d)$ is measured. The scaling
behaviour $q(d)\approx d^{-1/\nu_{3}}$ is well verified in this
case also, in accordance with eq.~(\ref{eq:10}).


\section{\label{pc2d3dfree}Percolation threshold in the direction parallel to a 
film of thickness $h$}

\begin{figure}
\includegraphics{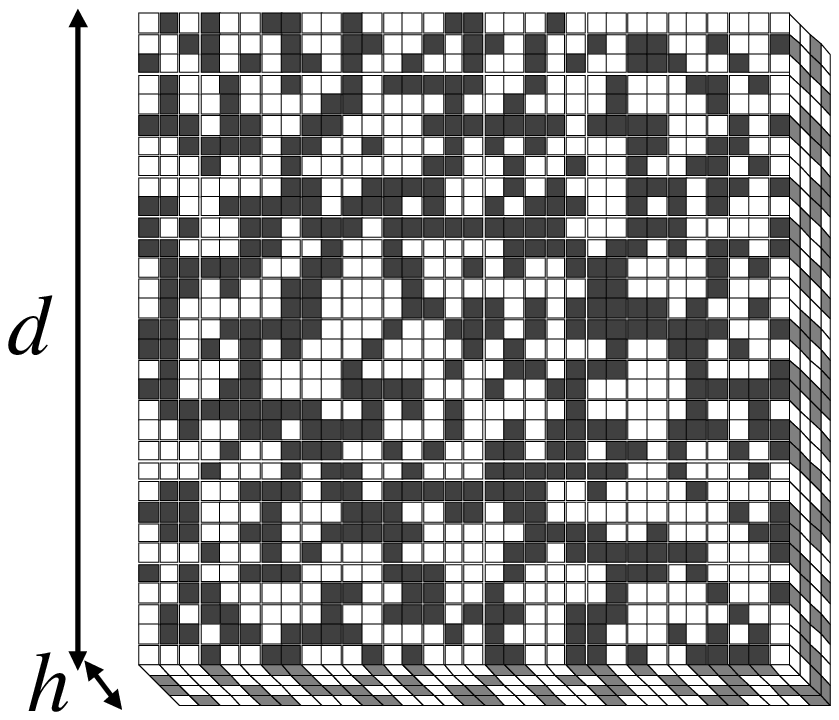}
\caption{\label{fig2} Percolation in a system of dimensions $d
\times d \times h$: there is a cluster spanning the whole surface
of the system. This figure illustrates the case of a non-adsorbing
film. }
\end{figure}

\begin{figure}
\includegraphics{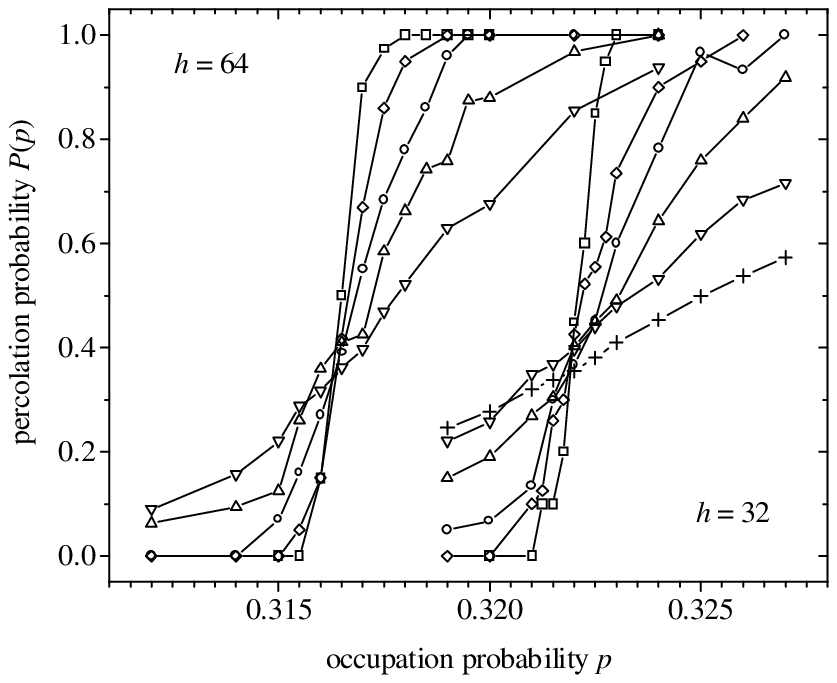}
\caption{\label{fig3} The probability to percolate $P$ as a
function of the occupation probability $p$, obtained in systems of
thickness $h$ ($h=32$ and $h=64$) and lateral extension $d$:
$\square : d=1024$, $\diamond : d=512$, $\circ : d=256$,
$\triangle : d=128$ ,$\triangledown : d=64$, $+ : d=32$. For each
value of $h$, the curves $P(p)$ become steeper as $d$ increases
and intersect at a fixed point $p_{c}(h)$.}
\end{figure}

\begin{figure}
\includegraphics{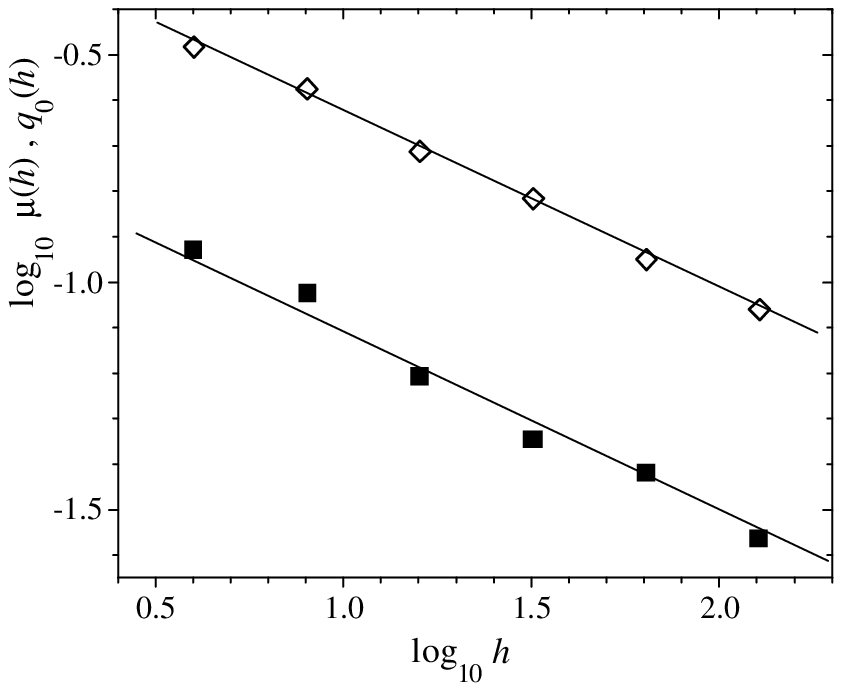}
\caption{\label{fig4} $\blacksquare$ : the prefactor $\mu (h)$ in
the curves $p_{c}(h,d)-p_{c}(h)$ versus $d$, as a function of $h$,
$\diamond$ : the prefactor $q_{0}(h)$ in the adjustable parameter
$q(d,h)$, in logarithmic scale. The straight lines show the
predicted behaviour $\mu(h), q_0(h) \propto h^{-0.386}$.}
\end{figure}

We consider the variation of the percolation threshold in a
system of dimensions $d \times d \times h$ as a function of the
thickness $h$ of the system. Note that the corresponding results are of 
interest for calculating the negative shift of $T_g$ in a 
freely suspended films \cite{long1,long3}. The thickness $h$ was varied from $h
= 1$ (which corresponds to the 2D system) to $h = 128$. For each
value of $h$, the dimension $d$ was varied from $d = h$ to $d =
1024$. The algorithm combines 3D renormalization up to a cube of
dimensions $h \times h \times h$, then 2D renormalization of
$h$-cubes up to a system of dimensions $d \times d \times h$.
Percolation is defined here when one cluster at least connects all
6 external faces of the volume $d \times d \times h$ (see
Fig.~\ref{fig2}). For each couple of values $(d,h)$, the
probability to percolate $P$ is measured as a function of the
occupation probability $p$. Some examples of the curves $P(p)$ are
shown in Fig.~\ref{fig3}. For each value of $h$, the curves
corresponding to different values of $d$ intersect at a fixed
point, which may be considered as the percolation threshold
$p_c(h)$. The curves are fitted by the same heuristic function as
before, with two adjustable parameters $p_{c}(d,h)$ and $q(d,h)$ :
\begin{equation}
P(p)=\frac{1}{2} \left( 1+\tanh \left(\frac{p-p_{c}(d,h)}{q (d,h)}
\right) \right)
\end{equation}
The number value giving the best linear regression
$p_{c}(d,h)-p_{c}(h) \approx d^{-1/\nu_{2}}$ ($h$ being fixed),
with the 2D value of the exponent $\nu_{2}=4/3$, will be adopted
for the percolation threshold $p_{c}(h)$ in a film of thickness
$h$, in the limit $d \rightarrow \infty$. Within error bars, this
value $p_{c}(h)$ coincides with the fixed point mentionned above in
the ensemble of curves $P(p)$. Thus, for a given thickness $h$ and
varying lateral size $d$, the following variation is observed:
\begin{equation}
p_{c}(d,h)-p_{c}(h) \approx \mu (h)d^{-1/\nu_{2}} \label{eq:24}
\end{equation}
with $\nu _{2}=4/3$. The values obtained for the percolation
thresholds $p_c(h)$ in a film of thickness $h$ are summarized in
Table \ref{tab:table2}.
\begin{table}
\caption{\label{tab:table2}The values of the percolation
thresholds in a film of thickness $h$}
\begin{ruledtabular}
\begin{tabular}{lr}
$h$&$p_{c}(h)$\\
\hline
2 & 0.47424\\
4 & 0.3997\\
8 & 0.3557\\
16 & 0.3329\\
32 & 0.3219\\
64 & 0.3165\\
128 & 0.31398\\
\end{tabular}
\end{ruledtabular}
\end{table}
The variation of the prefactor $\mu (h)$ may be estimated as
follows. The 3D behaviour (up to the size $h$) gives
$p_{c}(d)-p_{c}^{3D}=Ad^{-1/\nu_{3}}$ (see eq.~(\ref{eq:11})).
Thus, setting $d=h$ in eq.~(\ref{eq:24}):
\begin{equation}
p_{c}(h)+\mu (h)h^{-1/\nu_{2}} \approx p_{c}^{3D}+Ah^{-1/\nu_{3}}
\end{equation}
which gives, since $p_{c}^{3D}-p_{c}(h) \approx h^{-1/\nu_{3}}$:
\begin{equation}
\mu (h) \approx h^{1/\nu_{2}-1/\nu_{3}}\label{eq:25}
\end{equation}
The curve $\mu(h)$ is shown in Fig.~\ref{fig4} . The variation is
compatible with the exponent $1/\nu_{2}-1/\nu_{3}=-0.386$. Note
that the prefactor in eq.~(\ref{eq:25}) is arbitrary, since it
depends on the probability level $P$ which is adopted to define
percolation.

\begin{figure}
\includegraphics{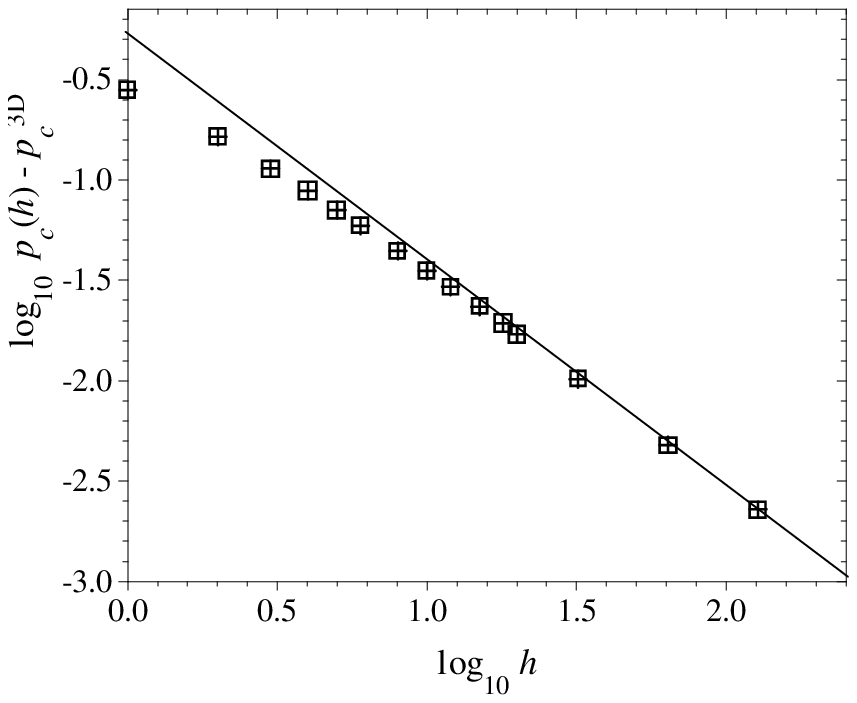}
\caption{\label{fig5} The reduced percolation threshold
$p_{c}(h)-p_{c}^{3D}$ as a function of the thickness $h$ of the
system, in logarithmic scale. The straight line represents the
scaling law with the 3D exponent $-1/\nu_{3}=-1.1364$. The scaling
law is observed asymptotically.}
\end{figure}

The same argument may be used to study the variation of the
fitting parameter $q(d,h)$. It is indeed observed that, varying
$d$ for a fixed value of $h$, a 2D scaling behaviour is observed:
$q(d,h)=q_{0}(h)d^{-1/\nu_{2}}$ (see Fig.~\ref{fig4}). The
prefactor is observed to vary as $q_{0}(h)= q_{0} h^{-0.386}$, in
accordance with the expected variation, which may be infered by
exactly the same argument as was used for $\mu(h)$
(eq.~(\ref{eq:25})). The prefactor $q_{0}$ is not arbitrary here:
it has the value 0.565.
In Fig.~\ref{fig5}, the percolation threshold $p_c(h) - p_c^{3D}$
is plotted as a function of $h$ in logarithmic scale. The best
linear fit has a slope $-1.1364$ which corresponds to the 3D value
$1/\nu_{3}=1/0.88$ of the exponent, in agreement with
eq.~(\ref{eq:13-6}). Thus, in the asymptotic limit, the curve
$p_c(h)$ writes :
\begin{equation}
p_{c}(h)-p_{c}^{3D} \approx  \alpha h^{-1.138}\label{eq:30}
\end{equation}
From Fig.~\ref{fig5}, the prefactor $\alpha$ has the value
$\alpha \approx 0.45$. Note that this value is of the same
order as, but different from the value
$p_{c}^{2D}-p_{c}^{3D}=0.281$ (cf eq.~(\ref{eq:13-5})).  

\section{\label{pc2d3dads} Correlation function between both interfaces 
of a film of thickness $h$}

\begin{figure}
\includegraphics{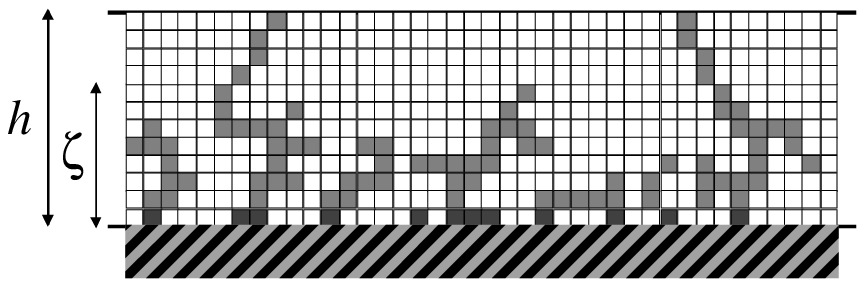}
\caption{\label{fig6} Schematics of a film of thickness $h$ in the
regime $\zeta < h$. Only percolation clusters connected to the
lower surface (i.e. to the substrate) are shown. Very few clusters
reach the upper surface. }
\end{figure}

We consider here the correlation function between both interfaces 
of a thin film. The regime of interest here is that of a thickness 
$h$ larger than the 3D correlation length of the percolation problem. 
Note that the corresponding results are of interest for calculating 
the positive shift of $T_g$ in a strongly adsorbed film \cite{long1,long3}. 
Indeed, in the case of a strongly interacting film, the glass transition 
is related to the
presence of \emph{one} percolating cluster connecting the upper
surface to the lower one (see Fig.~\ref{fig6}). In the sense of
percolation, glass transition will thus occur in this case, as
soon as the 3D correlation length $\zeta$ becomes of the order of (still being smaller than)  
of $h$. The effective percolation threshold defined in this way is
thus shifted downwards with respect to $p_{c}^{3D}$ (since
$p_{c}^{3D}$ corresponds to $\zeta \rightarrow \infty$).
Let us consider the average number of sites to which an occupied 
site of a given interface is connected to the other interface. 
This number $n_{c}(h)$ can be calculated by integrating the (3D)
correlation function, expressed in eq.~(\ref{eq:4g}):
\begin{eqnarray}
n_{c}(h) \approx 
\int_{0}^{\infty}\left(r^2+h^2\right)^{-\frac{1+\eta_3}{2}}
e^{-\frac{\left(r^2+h^2\right)^{1/2}}{\zeta}}rdr\nonumber \\
=\int_{h}^{\infty}\rho^{-\eta_3}e^{-\frac{\rho}{\zeta}}d\rho
\label{eq:32a}
\end{eqnarray}
Integrating by parts, this gives:
\begin{equation}
n_{c}(h) \approx h^{-\eta_3} \zeta e^{-\frac{h}{\zeta}} -\eta_3\zeta
\int_{h}^{\infty}\rho^{-1-\eta_3}e^{-\frac{\rho}{\zeta}}d\rho
\label{eq:32b}
\end{equation}
The second term is negligible for $h\gg\zeta$.

In our simulations we have calculated the number $n_s(h)$ of sites of 
one interface 
which are connected to the other interface with at least one site. We expect 
a similar dependence of $n_c(h)$ as that given for $n_s(h)$ by equation 
eq.~(\ref{eq:32b}). In our simulations, 
we consider a system of dimension $d \times d \times h$ (with $d$
large with respect to $h$). For a given value of the occupation
probability $p$ such that $\zeta < h$, we compute 
the number $n_{s}(h)$. The value of $d$ is $d
= 1024$, $h$ ranges from $h = 1$ to $h = 64$. The algorithm makes
use of a 1D renormalisation procedure : 2D slices of size $d
\times d$ are generated and then glued together recursively,
leading to systems of thickness $2$, then $2^2$, etc, up to $2^6 =
64$.

For a given value of the occupation probability $p$ (with
$p<p_{c}^{3D}$), the quantity $\log n_{s}(h)$ exhibits a linear
behaviour in the limit of large $h$ values. Each curve was fitted
with the first term in eq.~(\ref{eq:32b}), with two independent
adjustable parameters $\eta_3$ and $\zeta$. The values obtained for
$\zeta$ in this way differ by less than $15\%$ from those obtained
by simply taking the slope of the asymptotic linear variation at
large $h$ values, i.e. by neglecting the influence of the
$h^{-\eta_3}$ factor (remember that the exponent $\eta$ is quite
small). This difference is systematic and does not alter the 
variation as a function of $p_c^{3D}-p$ as it is 
shown in Fig. 7.

\begin{figure}
\includegraphics{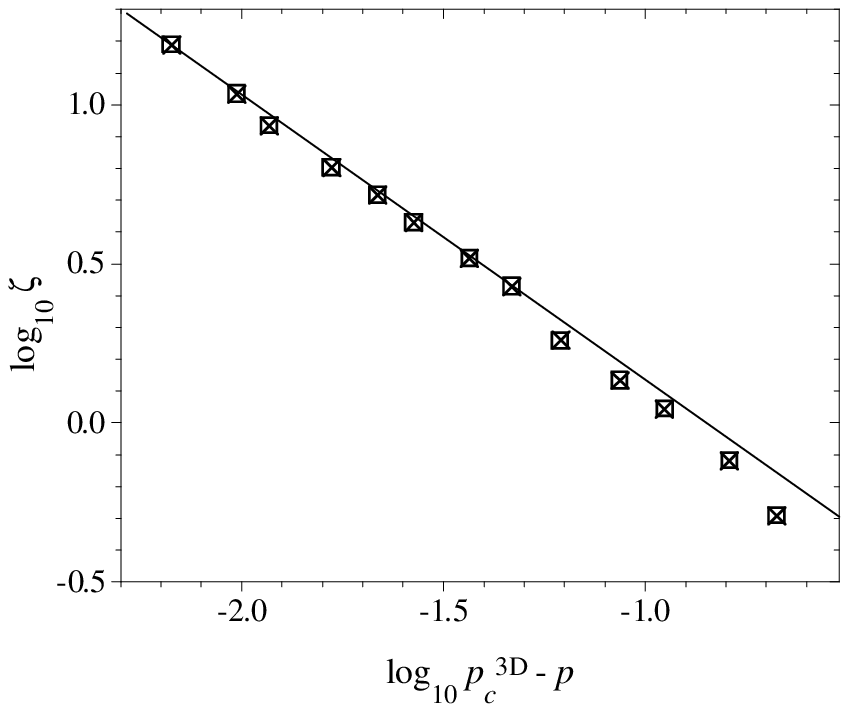}
\caption{\label{fig7} The correlation length $\zeta(p)$ obtained
from a linear fit of the curves $n_{s}(h)$ at large $h$ values
(see eq.~(\ref{eq:32b})), versus $p_{c}^{3D}-p$, in logarithmic
scale ($p_{c}^{3D}$ is the 3D percolation threshold $p_{c}^{3D} =
0.3117$). }
\end{figure}

In Fig.~\ref{fig7}, the correlation length $\zeta(p)$ is plotted
as a function of $p_{c}^{3D} - p$, in logarithmic scale. The data
are compared to a straight line of slope $-\nu_{3}=-0.88$, which
is the behaviour expected in the limit $p \rightarrow p_{c}$. An
excellent agreement is obtained in the asymptotic limit. Thus we
have:
\begin{equation}
\zeta (p) \approx \zeta_{0} \left( p_{c}^{3D}-p
\right)^{-\nu_{3}}\label{eq:34}
\end{equation}
with $\zeta_{0} \approx 0.185$. Note that two opposite requirements have
to be reconciled in this section: observing the critical power law
$\zeta(p) \approx \left(p_{c}-p \right)^{-\nu_{3}}$ urges for
coming close enough to $p_c^{3D}$, while maintaining $\zeta$
small enough to make the regime $h>\zeta$ observable. 

\section{\label{libreads}Cluster number in a film of thickness $h$}
\begin{figure}
\includegraphics{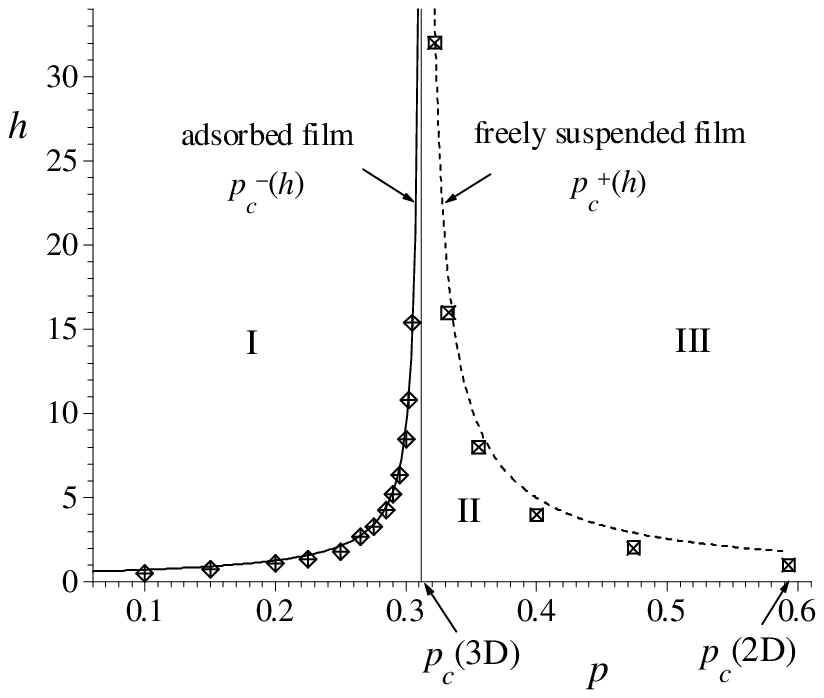}
\caption{\label{fig8} Diagram of the percolation thresholds
$p_{c}^{-}(h)$ and $p_{c}^{+}(h)$ as a function of the film
thickness $h$.
 $p_{c}^{-}(h)$ is the percolation threshold in an adsorbed film.
It is defined here as the ensemble of points for
which $\zeta ^{3D}(p) \equiv h$ . Diamonds symbols correspond to
the same data set as in Fig.~\ref{fig7}. The continuous line is
the fit with equation~\ref{eq:37}. $p_{c}^{+}(h)$ is the
percolation threshold in a freely suspended film. Square symbols
corresponds to the data in Fig~\ref{fig5}. The dashed line is the
fit with eq.~(\ref{eq:30}) in section ~\ref{pc2d3dfree}.  }
\end{figure}

\begin{figure}
\includegraphics{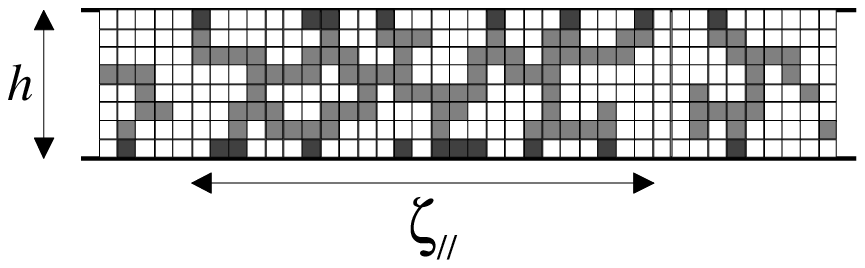}
\caption{\label{fig9} Schematics of a cluster percolating through
the film thickness, in the regime $h < \zeta <d$. The 'surface
mass' $s$ of the cluster is given by the ensemble of sites at the
surface of the film, which are emphasized. }
\end{figure}

The regime of interest here corresponds to a 3D correlation length 
larger than the thickness of the film, either below the 3D percolation threshold, or above.  
The diagram in Fig.~\ref{fig8}
summarizes the results obtained so far. 
$p$ is the occupation probability, $h$
the film thickness (that is, the number of slices in the simulated
system). The curve $p_{c}^{-}(h)$ is defined as the ensemble of
points where the measured 3D correlation length $\zeta$ is equal
to $h$. Note that this curve corresponds to the percolation threshold
associated to glass transition in a strongly adsorbed film, as it
was defined in Section~\ref{pc2d3dads}. According to
eq.~\ref{eq:34}, it is described by the equation:
\begin{equation}
p_{c}^{3D}-p = \left( \frac{h}{\zeta_{0}}
\right)^{-1/\nu_{3}}\label{eq:37}
\end{equation}
The curve $p_{c}^{+}(h)$ is defined by eq.~\ref{eq:30}. It
corresponds to the percolation threshold in the direction 
parallel to the film. Note that it corresponds 
also to the glass transition in a freely suspended
film of thickness $h$, as defined in Section~\ref{pc2d3dfree}.
The two curves $p_{c}^{-}(h)$ and $p_{c}^{+}(h)$ define 3 regions
in the plane $(p,h)$. In region I, one has $\zeta<h<d$. Clusters
connecting both faces of a film of thickness $h$ are exponentially
rare. In Region III,
there is an infinite cluster extending throughout a film of
thickness $h$. Region II corresponds to
intermediate cases $h<\zeta<d$. 
For a given thickness $h$, this corresponds to going from $p_{c}^{+}(h)$ to
$p_{c}^{-}(h)$ within region II. In this region, some clusters 
have a lateral extension large with respect to $h$. As one comes closer
to the curve $p_{c}^{+}(h)$, the correlation length parallel to
the plane of the film, i.e. the lateral extension of clusters
$\zeta$ diverges. One can see such an aggregate in Fig.~\ref{fig9}.
Thus, in this section, we study the distribution of cluster
sizes and masses, both on the surface and in the bulk of a film of
thickness $h$. The corresponding results are of interest for 
describing the cross-over between freely suspended films 
and strongly adsorbed films as far as glass 
transition is concerned \cite{long3}. 

We will check that the distribution
of clusters and their structure may be described by the procedure
detailed in Section~\ref{rescale}, in which we assumed that a
percolation system of thickness $h$ may be renormalized to a true
2D system by coarse-graining at the scale $h$.
Note that the system has to be larger than the typical lateral
extension of large clusters, which become larger as one comes
closer to $p_{c}^{+}(h)$. On the other hand, the second moment of
the distribution of cluster masses $n(s)$ is dominated by the
contribution from large clusters, which are exponentially rare.
Thus, we need here to simulate both large enough systems and a
large enough number of realisations of a given system, which is
quite demanding in terms of computer size and time.
The volume $V=h \times d^2$ of the simulated systems is of the
order a few $10^8$ typically. Results are
averaged over 20 to 50 different realisations, depending on the
distance to the $p_{c}^{+}(h)$ curve.

\subsection{\label{masvol}The average mass of the clusters}
Let us consider a cluster in a film of thickness $h$, in the
regime $h < \zeta < d$ (i.e. in Region II of the diagram in
Fig.~\ref{fig8}). It was shown in Section~\ref{rescale} how a film
of thickness $h$ may be renormalized to a 2D percolation system.
Specifically, in the regime $h < \zeta < d$, the average mass $M$,
expressed in units of elementary sites and defined as in
eq.~(\ref{eq:1}), is given by eq.~(\ref{eq:15bis}), i.e.:
\begin{equation}
M(h,p) =M_{0} h^{2-\eta_3} \left( h^{1/\nu_{3}} \vert p-p_{c}(h) \vert
\right)^{-\gamma_{2}}
 \label{eq:41}
\end{equation}
The 2D behaviour is reflected in the exponent $-\gamma_2$. We have
obtained the distribution of cluster masses $n(m)$ for different
values of $p$ and $h$ (within region II) in systems of total
volume of the order $2.6 \times 10^8$, and computed the second
moment $M(h,p)$ (see eq.~(\ref{eq:1})).

\begin{figure}
\includegraphics{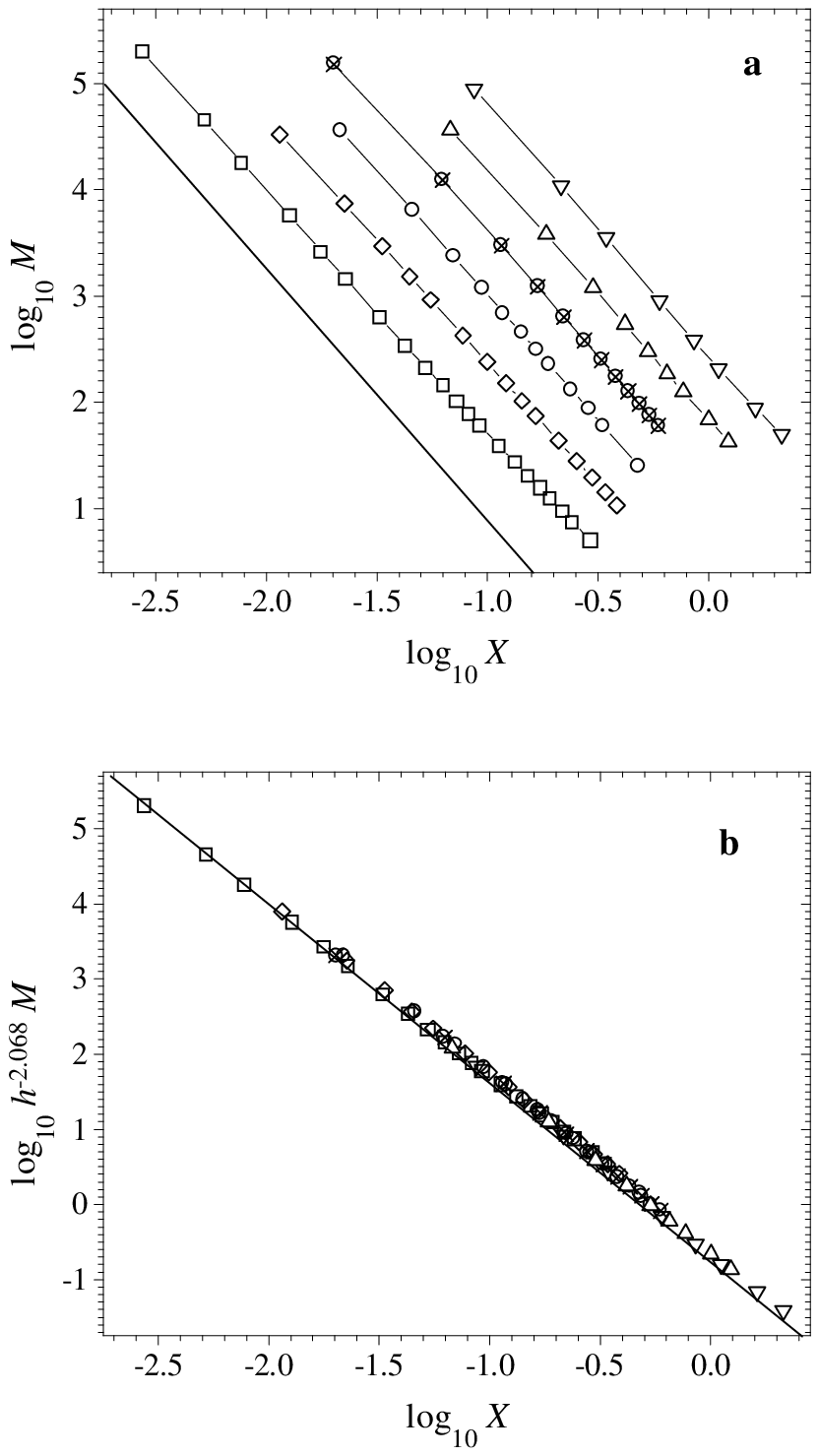}
\caption{\label{fig10} {\bf a}: the average mass $M(h,p)$ vs
 $X=h^{1/\nu_{3}}\left( p_{c}(h)-p
\right)$, {\bf b}: the quantity $M'=h^{-2+\eta_{3}}M(h,p)$ vs
 $X$, for different values of $h$: $\square : h=1$, $\diamond :
h=2$, $\circ : h=4$, $\otimes : h=8$, $\triangle : h=16$,
$\triangledown : h=32$. The straight lines corresponds to the
exponent $-\gamma_{2} =43/18=2.389$. In {\bf b}, a master curve is
obtained to an excellent approximation.}
\end{figure}

Let us first consider the variation versus $p$, $h$ being fixed.
For each value of $h$, the variation of $M(h,p)$ versus $p_{c}(h) -p$ 
gives a power law with the exponent
$-\gamma_{2}=-2.389$ in the limit $p$ close to $p_{c}(h)$, in
agreement with the expected behaviour. To illustrate this power
law and the influence of the thickness $h$ of the system, the
curves $M(h,p)$ are plotted as a function of the quantity
$X=h^{1/\nu_{3}}( p_{c}(h) -p)$ in Fig.~\ref{fig10}a
for different values of $h$. The values at $\log X=0$ for each
portion of curve in Fig.~\ref{fig10} (corresponding to a given
value of $h$) gives the prefactor $M_{0}(h)$ in the curves
$M(p,h)$ versus $X$. In the limit of large $h$, the behaviour of
the prefactor $M_{0}(h)$ is compatible with a power law
$M_{0}(h)=M_{0}h^{2-\eta_{3}}$ ($2-\eta_{3}=2.068$). The prefactor
$M_0$ has the value $M_{0} \approx 0.233$.

The quantity $M'=h^{-2+\eta_{3}}M(h,p)$ plotted versus $X$ should
then give a master curve of the form $M'=M_{0}'X^{-\gamma_{2}}$.
This is illustrated in Fig.~\ref{fig10}b. The expected behaviour
is indeed observed to an excellent approximation in a large range
of variation (more than 4 decades in $M$). The full scaling
behaviour of $M(h,p)$ as a function of $p_{c}(h)-p$ and $h$ (eq. (57)) is
thus summarized in Fig.~\ref{fig10}b. Remember that $M'$
represents the mass expressed in number of renormalized
supersites.
Alternatively, the same work may be done with the
surface mass $s$ of the clusters, i.e. the number of sites
belonging to a cluster which are in contact with the substrate
(the surface mass of clusters is denoted by $s$ in order to
distinguish it from their total mass $m$. See Fig.~\ref{fig9}).
The average surface mass was obtained in the following way. The
distribution $n(s)$, which gives the number of clusters with a
surface mass $s$, was recorded for different values of $h$ and $p$
(in region II of the diagram in Fig.~\ref{fig8}). The average
surface mass $S(p,h)$ was then computed as the second moment:
\begin{equation}
S(h,p)  = \frac{\sum s^{2}n(s)}{\sum s\,n(s)} \label{eq:50}
\end{equation}
where the summation extends over all clusters which percolate
through the film thickness.

\begin{figure}
\includegraphics{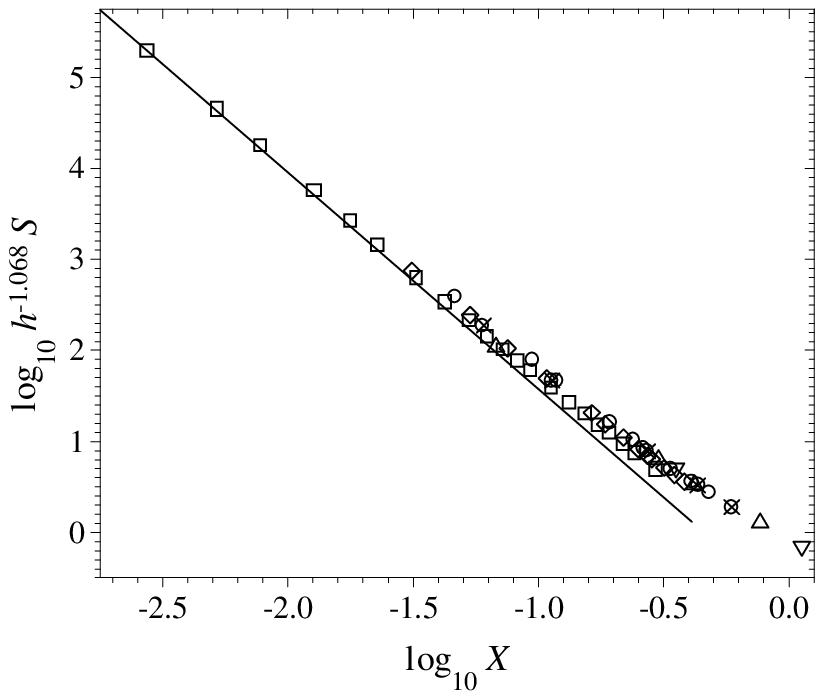}
\caption{\label{fig11} The reduced average surface mass
$h^{-1+\eta_3}S(h,p)$ vs the quantity $X=h^{1/\nu_{3}}\left( p_{c}(h)-p
\right)$ in log scale, for different values of the thickness $h$.
The straight line corresponds to the exponent $-\gamma_{2} =
43/18=2.389$.}
\end{figure}

In a way similar to $M(h,p)$, $S(h,p)$ is expected to scale like:
\begin{equation}
S(h,p) =S_{0} h^{1-\eta_3} \left( h^{1/\nu_{3}} \vert p-p_{c}(h) \vert
\right)^{-\gamma_{2}}
 \label{eq:51}
\end{equation}
When plotted versus $\left( p-p_{c}(h) \right)$, for each value of
$h$, $S(h,p)$ gives a power law with the exponent
$-\gamma_{2}=-2.389$, in agreement with the expected behaviour and
with the behaviour observed for the quantity $M(p,h)$. From
eq.~(\ref{eq:51}), a master curve should be obtained when the
quantity $h^{\eta_3-1}S(h,p)$ is plotted versus
$X=h^{1/\nu_{3}}\left( p_{c}(h)-p \right)$. This is shown in
Fig.~\ref{fig11}. Here also, the expected asymptotic behaviour is
observed to an excellent approximation in a large range of
variation for $S$. The prefactor $S_0$ has the value $S_{0}
\approx 0.24$.

\subsection{\label{distrmas}The distribution of cluster masses in the surface of the film}
\begin{figure}
\includegraphics{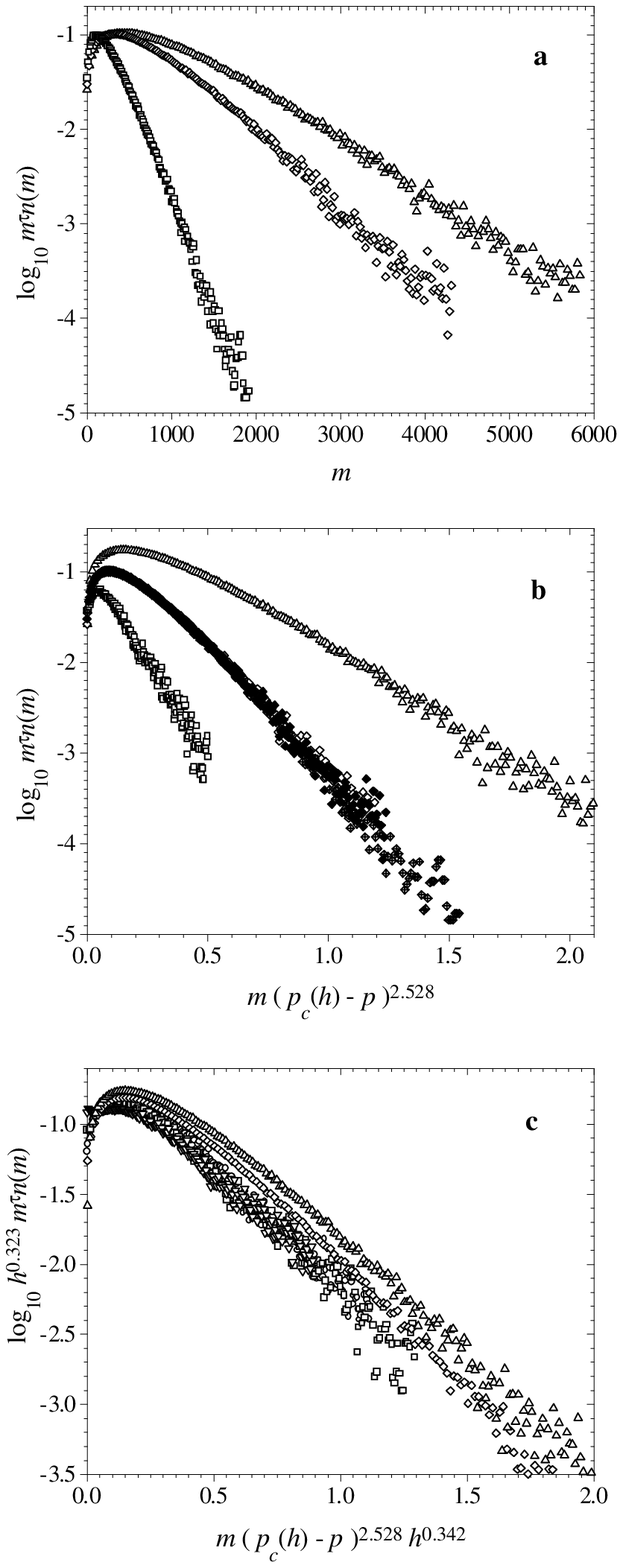}
\caption{\label{fig12}  Scaling properties of the distribution
$m^{\tau_2}n(m)$, where $n(m)$ is the number of clusters of mass
$m$ per lattice site . a: $\log m^{\tau}n(m)$ plotted versus $m$
for a thickness $h=4$ and different values of the occupation
probability $p$: $\square : p=0.34$, $\diamond$ : $p=0.36$,
$\triangle : p=0.365$. b: $\log m^{\tau}n(m)$ plotted versus
$\vert p-p_{c}(h)\vert^{1/\sigma_{2}} m$, for different values of
$h$ : $\triangle$ : $ h=1$, $\diamond$ and 
$\blacklozenge$ : $h=4$,
$\square$ : $h=16$. In this representation, the curves obtained for
different values of $p$ at fixed $h$ superpose. c:
$h^{-\omega}\log m^{\tau}n(m)$ plotted versus the quantity
$h^{0.342}\vert p-p_{c}(h)\vert^{1/\sigma_{2}} m$, for different
values of $p$ and $h$: $\triangle : p=0.54,h=1$, $\diamond :
p=0.34,h=4$, $\circ : p=0.34,h=8$, $\square : p=0.32,h=16$,
$\triangledown : p=0.31,h=32$. The curves superpose in the
asymptotic limit of large values of $h$ and $p_{c}(h)-p$. }
\end{figure}

In this Section, we show that not only the average mass, but also
the whole distribution of cluster masses obeys scaling laws in the
two variables $h$ and $p-p_{c}(h)$, which allows to draw a master
curve. According to Eq.~(\ref{eq:16-0}), the number of clusters of
mass $m$ obeys (in the limit of large $m$):
\begin{equation}
h^{-\omega}m^{\tau_{2}}n(m) \approx g_{2D} \left[
h^{-D_3+1/\sigma_{2}\nu_{3}}\vert p-p_{c}(h)\vert^{1/\sigma_{2}}
m\right]
 \label{eq:61}
\end{equation}
i.e., the quantity $h^{-\omega}m^{\tau_{2}}n(m)$ should give a
master curve when plotted versus the quantity
$h^{-D_3+1/\sigma_{2}\nu_{3}}\vert p-p_{c}(h)\vert^{1/\sigma_{2}}
m$. The characteristic exponents of $h$ have the values
$-D_3+1/\sigma_{2}\nu_{3}=0.342$ and $\omega=-0.323$. 
Consider first the variation versus $p-p_{c}(h)$, $h$ being fixed.
The distributions of clusters, represented by the quantity
$m^{\tau_{2}}n(m)$, obtained for $h=4$ and different values of
$p-p_{c}(h)$, are plotted as a function of $m$ in
Fig.~\ref{fig12}a. For each value of $h$ ($h=1$ to $h=32$), an
excellent superposition property is observed when
$m^{\tau_{2}}n(m)$ is plotted as a function of the reduced
variable $\vert p-p_{c}(h)\vert^{1/\sigma_{2}}m$. This is
illustrated in Fig.~\ref{fig12}b, in which the various curves
shown in Fig.~\ref{fig12}a ($h=4$) are superposed.

Two exponents have to be adjusted in order to superpose the
ensembles of curves obtained for different values of $h$. First,
plotting the quantity $\log m^{\tau_{2}}n(m)$ as a function of the
reduced variable $h^{-D_{3}+1/\sigma_{2}\nu_{3}}\vert
p-p_{c}(h)\vert^{1/\sigma_{2}}m$ allows the slopes obtained in the
asymptotic limit at large $m$ to coincide. Then, the quantity
$\log h^{-\omega} m^{\tau_{2}}n(m)$ is plotted as a function of the
reduced variable $h^{-D_3+1/\sigma_{2}\nu_{3}}\vert
p-p_{c}(h)\vert^{1/\sigma_{2}}m$.  This procedure is illustrated
in Fig.~\ref{fig12}c. As mentionned above, according to
eq.~\ref{eq:61}, a master curve should be obtained.

The quantity $\log h^{-\omega} m^{\tau_{2}}n(m)$ does indeed show a
linear asymptotic behaviour at large $m$ values, i.e. for clusters
larger than the correlation length $\zeta$. It is observed in
Fig.~\ref{fig12}c that the slopes obtained at large $m$ values
indeed coincide, within numerical uncertainties. In contrast,
vertical superposition of the curves obtained for various $h$
values is only observed asymptotically, i.e. towards the largest
investigated value $h=32$. The large dispersion in the curves
comes from the numerical limitations of the present simulations,
especially for large values of $h$ and/or $p_{c}(h)-p$. In this
regime, where the correlation length $\zeta$ is large, the lateral
size $d$ of the system has to be large, and the distributions
$n(m)$ extend towards large $m$ values. As already mentionned,
computing the distributions $n(m)$ with a good sampling of large
$m$ values is very demanding in terms of memory size and computer
time. In any case, it can be concluded at this point that our
data are compatible with the scaling law in eq.~\ref{eq:61} in the
asymptotic regime of large $h$ values, which is particularly clear when 
considering the data corresponding to the thicknesses $h=8,16$ 
and $32$ respectively. 

\begin{figure}
\includegraphics{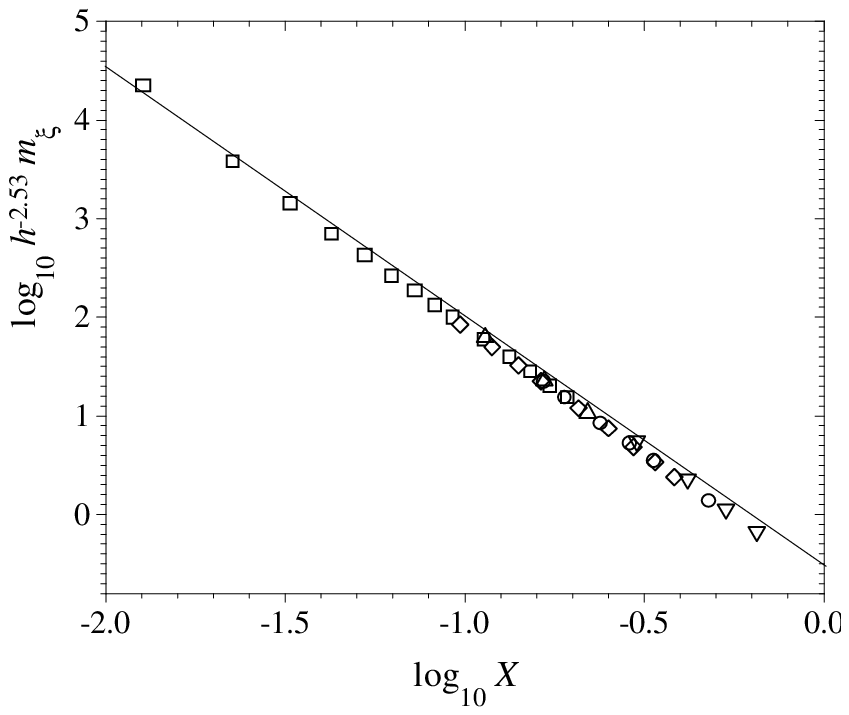}
\caption{\label{fig13} The quantity $m_{\zeta}$ obtained by
fitting the distribution $m^{\tau_2}n(m)$ at large $m$. The
quantity $h^{-D_3}m_{\zeta}$ is plotted versus
$X=h^{1/\nu_{3}}\left( p-p_{c}(h)\right)$, for different values of
$h$: $\square : h=1$, $\diamond : h=2$, $\circ : h=4$, $\triangle
: h=8$, $\triangledown : h=16$. A master curve is obtained to an
excellent approximation. The straight line corresponds to the
exponent $-1/\sigma_{2} =-2.528$. }
\end{figure}

 To illustrate further the scaling properties expressed in eq.(~\ref{eq:61}),
we have fitted the exponential decrease of $m^{\tau_{2}}n(m)$ at
large $m$ values. This procedure gives the quantity
$m_{\zeta}(p,h)$, which depends on $p$ and $h$ and may be
considered also as a typical mass describing the distribution of
the clusters within the film (see eq.(~\ref{eq:2})). From
eq.(~\ref{eq:61}), $m_{\zeta}$ obeys the following scaling law:
\begin{equation}
m_{\zeta} \approx  h^{D_3} \left[h^{1/\nu_{3}}\vert
p-p_{c}(h)\vert \right]^{-1/\sigma_{2}}
 \label{eq:61bis}
\end{equation}
First, for each value of $h$, it is observed that $m_{\zeta}(p,h)$
follows a scaling law when plotted as a function of $\left(
p-p_{c}(h) \right)$, with the exponent $-1/\sigma_{2}=-2.528$.
Then a master curve is obtained when $h^{-D_3}m_{\zeta}(p,h)$ is
plotted as a function of the quantity $h^{1/\nu_{3}}\left(
p-p_{c}(h)\right)$. The obtained master curve is shown in
Fig.~\ref{fig13}.

\subsection{\label{struct}The structure of the clusters in a film of thickness $h$}
The structure of the clusters is described by the relationship
between the mass $m$ of a cluster (the number of sites belonging
to the cluster) and its size $r$. Clusters have a fractal
structure characterized by the exponent $D$ defined as $r \approx
m^{1/D}$, in which the value of $D$ effectively measured varies
according to the location of the system with respect to $p_c$, or
equivalently, according to the typical range of sizes which is
probed compared to the correlation length $\zeta(p)$ (see
eq.~(\ref{eq:15}) and the discussion thereafter). The cluster size
may be defined as either the gyration radius or some measure of
the cluster extension.

\begin{figure}
\includegraphics{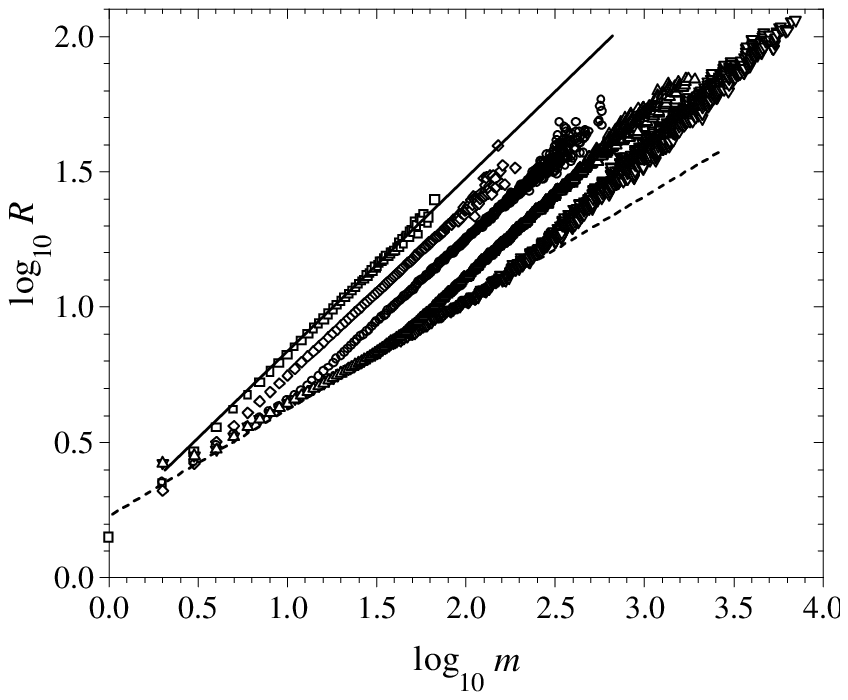}
\caption{\label{fig14} The average lateral size $R= \left(
\overline{\Delta x^{2}+\Delta Y^{2}}\right)^{1/2}$ is plotted as a
function of the total mass $m$ of the clusters, in films of
various thickness $h$: $\square : h=1$, $\diamond : h=2$, $\circ :
h=4$, $\triangle : h=8$, $\triangledown : h=16$. The occupation
probability $p = 0.315$. The straight lines have a slope 1/1.561,
which corresponds to the fractal exponent $D_{2} = 1.56$ at
$p<p_c$.
 }
\end{figure}

We have studied the relationship between the size and mass of
clusters in a film of thickness $h$, within the interval
$p_{c}^{-}(h)<p<p_{c}^{+}(h)$. The squared size of a cluster $i$
is defined here as $r_{i}^{2}=\Delta x_{i}^{2}+\Delta y_{i}^{2}$ ,
where $\Delta x_i$ and $\Delta y_i$ are the maximum extensions of
the cluster $i$ along $x$ and $y$ ($x$ and $y$ are the coordinates
parallel to the plane of the film), or more precisely, the maximum
extensions of the intersection of the cluster with the limiting
surfaces of the film. The mass $m_i$ is the total number of sites
which belong to cluster $i$ (within the whole volume of the film).
The quantity $r_{i}^{2}$ is then averaged over all clusters of
mass $m$, in order to obtain the squared average size
$\overline{r^{2}}(m)$ as a function of $m$:
\begin{equation}
\overline{r^{2}}(m)=\frac{1}{Vn(m)} \sum _{i} r_{i}^{2}
\end{equation}
where the sum is extended over the $Vn(m)$ clusters having the
mass $m$ ($V=d^{2} \times h$ is the volume of the system). The
relationship between the mass $m$ and the average size
$R=(\overline{r^{2}}(m))^{1/2}$ of the clusters in a film of
variable thickness $h$ is illustrated in Fig.~\ref{fig14}.

Some examples of the curves $R(m)$, obtained in a system of
lateral dimension $d=2048$ and different values of the thickness
$h$ (from $h = 1$ to $h = 16$) are plotted in logarithmic scale in
Fig.~\ref{fig14}, for an occupation probability $p = 0.315$. 
The observations which can be made on Fig.~\ref{fig14} are the following. 
First, two different regimes are observed. They are
characterised by well defined power laws, with different
exponents. At small $R$ values (typically, $R$ smaller than or of
the order of $h$), the mass grows with $R$ faster than for larger
$R$ values. Indeed, clusters should behave in a 3D way in this
regime. Within the whole interval $p_{c}^{-}(h)<p<p_{c}^{+}(h)$,
the 3D correlation length is larger than $h$. Thus, the expected
behaviour in this regime should be described by the fractal
exponent $1/D_3$ \emph{at} $p_c^{3D}$, according to
eq.~(\ref{eq:15}) and the discussion thereafter. This is indeed
the case: the slope of this part of the curves is compatible with
the value $1/D_{3}=0.395$ at $p=p_{c}^{3D}$. This regime is
illustrated by the dashed line in Fig.~\ref{fig14}. Then, at $R
\approx h$, each curve has a crossover to another regime. For $R >
h$, one enters a quasi 2D regime, in which the correlation length
in the equivalent (renormalized) 2D system is small. Therefore,
the slope in this regime has to be compared to the value of the
fractal exponent at $p<p_{c}^{2D}$: $1/D_{2}^{<}=0.641$
\cite{stauffer}. This quasi 2D regime is illustrated by the solid
line in Fig.~\ref{fig14}. The observed behaviour is indeed
compatible with the exponent $1/D_{2}^{<}$.

Thus, the results in Fig.~\ref{fig14} illustrates the quasi 2D
behaviour of the film: a film of thickness $h$ behaves as a
strictly 2D film. In the equivalent renormalized 2D
system, the 2D scaling law, eq.~\ref{eq:5},writes:
\begin{equation}
m'=m_{0}R'^{D_{2}^{<}}\label{eq:38}
\end{equation}
where $m_0$ is a prefactor of order unity (the superscript $<$
indicates that the exponent $D_2^{<}$ is that at $p<p_{c}^{2D}$).
Coming back to the original system of thickness $h$, the size
(expressed in units of elementary sites) becomes $R' \rightarrow
R=hR'$ and the mass is transformed as $m' \rightarrow
m=h^{D_{3}}m'$, since $\zeta_{3D}>h$. Eq.~(\ref{eq:38}) thus leads
to the following scaling:
\begin{equation}
m=m_{0}h^{D_{3}} \left(
\frac{R}{h}\right)^{D_{2}^{<}}\label{eq:39}
\end{equation}
From the data in Fig.~\ref{fig14}, the value of the prefactor
$m_{0} \approx 0.22$. Close inspection of the data in
Fig.~\ref{fig14} shows however that the apparent exponent in the regime
$R>h$ is slightly smaller than $1/D_{2}^{<}$ in some range of $R$
values at the beginning of this regime. Indeed, the scaling $m'
\approx R'^{D_{2}^{<}}$ is only valid when $R >\zeta_{\|}$,
where $\zeta_{\|}$ is the correlation length parallel to the film. 
In the range $h < R < \zeta_{\|}$, one should 
observe a power law $m \propto R^{1/D_2}$ where $1/D_2 = 49/91 \approx 0.538$ is the critical 
exponent at $p_c$. It just happens here that the range is too small for this regime 
to be clearly visible. For instance, 
$\zeta_{\|}$ (expressed in units of
$h$-cubes) is:
\begin{equation}
\zeta_{\|} \approx \zeta_{0} \left( h^{1/\nu_{3}} \vert
p_{c}^{+}(h)-p \vert \right)^{-\nu_{2}}\label{eq:40}
\end{equation}
which gives for the particular case of the data shown in
Fig.~\ref{fig14}, a value $\zeta_{\|} \approx 3.2$ (with a prefactor
$\zeta_0$ of the order 1). 

\section{\label{disc}Conclusion}
We have described here the cross-over between 2D and 3D percolation. 
This work demonstrates that a system of finite thickness $h$ may
be renormalized to a strictly 2D system for describing the percolation
properties, provided that the 2D percolation threshold
$p_{c}^{2D}$ is renormalized to an appropriate value
$p_{c}(h)$. The values $p_{c}(h)$ have been determined
numerically with a good precision for $h=2$ to $h=128$ (see Table
~\ref{tab:table2}), in the case of a cubic lattice.  
We have studied 3 different regimes, which describe 3 different
physical situations. These regimes are represented in
Fig.~\ref{fig8}. At a given value of the occupation probability 
$p$, if the thickness $h$ is larger than the 3D correlation length, 
the problem is a true 3D problem. The fraction of sites of one 
interface connected to the other interface by a cluster is 
exponentially small. The corresponding correlation function and 
correlation length are given by 3D critical exponents. Our 
numerical simulations provide the corresponding 
pre-factors. If, at a given value of $p$, the thickness 
$h$ is smaller than the 3D correlation length, then 
the problem is equivalent to a 2D problem. We provide 
the corresponding transformations for calculating 
the correlation length parallel to the film, the average mass 
of the clusters and the cluster mass distribution, among other 
relations. These quantities can be expressed 
as functions of known 2D and 3D critical exponents. Our numerical 
simulations provide the corresponding pre-factors for these 
quantities, as well as the master function for the mass distribution. 
These results can be used for describing the glass transition in thin 
polymer films as described in \cite{long3}, and should be also of interest 
for describing mechanical or electrical properties of 
films made of composite and disordered materials.

\end{document}